\documentclass[aps,prd,preprint,showpacs]{revtex4}
\usepackage{amssymb,amsmath,graphicx}

\newcommand{\be}{\begin{equation}}
\newcommand{\ee}{\end{equation}}
\newcommand{\ba}{\begin{eqnarray}}
\newcommand{\ea}{\end{eqnarray}}

\newcommand{\dcom}[1]{}
\newcommand{\dnote}[1]{}

\newcommand{\gsim}{\raise.3ex\hbox{$>$\kern-.75em\lower1ex\hbox{$\sim$}}}
\newcommand{\lsim}{\raise.3ex\hbox{$<$\kern-.75em\lower1ex\hbox{$\sim$}}}

\begin{document}
\title{Chaotic Motion of Charged Particles around a Weakly Magnetized Kerr-Newman Black Hole}
\author{Chen-Yu Liu}
\email{ef850502@gmail.com}
\affiliation{Department of Physics, National Dong Hwa University,
Hualien, Taiwan, R.O.C.}

\date{\today}

\begin{abstract}
In this paper, we consider a charged particle moves around a Weakly Magnetized Kerr-Newman black hole. We first study its circular motion with a detailed analysis in the innermost stable circular orbits(ISCO). Then the dynamics of a particle, kicked out from the circular orbit, is also explored. The final states can be characterized into three different types: bound motion, captured by black hole or escape to spatial infinity. The respective trajectories are shown, and are mainly determined by the energy of a particle after a kick and the initial condition, in which we use Basins of attraction to represent the relation between different initial condition and final states.
\end{abstract}


\maketitle

\section{Introduction}

Black hole plays an important role in the formation of galaxies~\cite{MJREE},they transfer potential energy of accretion disk to radiate energy~\cite{NISHA}.Recent result of observation show that the energy of the jet is proportional to black hole spin, which match Blandford-Znajek process~\cite{RNARA,RNARATC}. Systems as rotating black hole and accretion disk around it have some simplified researches, their main work is analysis the relation between black hole spin and I.S.C.O. of charged particle, the result is the same/different direction of rotating black hole and orbiting particle will lead to smaller/larger ISCO radius. Then give a kick on the particle initially in ISCO and see how initial conditions such as initial position of particle and the energy of the kick affect final states of particle by numerical method. In real case, kick energy may come from another particle or photon~\cite{RYOS,AMA1}. Similar research in Schwarzschild black hole background is also explored~\cite{ZAH}. Observation gives no evidence of the existence of charged black hole, one consider that a charged black hole will neutralized by plasma around it~\cite{GRA}.Before the neutralization, there are some theoretical research describe how black hole charge affect the ISCO of the neutral/charged particle around it~\cite{RUF1,RUF2}. Early days about the charged rotating black hole has born, we can describe it by Kerr-Newman metric. Equatorial plane of a neutral particle around it has researched by~\cite{RUF3}, when we give particle a kick like~\cite{RYOS,AMA1}, how initial conditions affect the final state will be explored in this paper. 
\section{Circular motion of a charged particle around Kerr Newman black hole}

The black hole geometry with a gravitational mass $M$, charge $Q$, and  angular momentum per unit mass $a=J/M$ is described
 by Kerr-Newman metric:
\ba
 {ds}^2 &=& g_{\mu\nu} dx^\mu dx^\nu  \nonumber \\
&= & -\frac{ \left(\Delta -a^2 \sin^2\theta \right)}{\Sigma } {dt}^2   + \frac{ a   \sin ^2\theta \left(Q^2-2 M
   r\right)}{\Sigma } ({dt}{d\phi+ d\phi dt)}  \nonumber\\
&&\quad\quad+  \frac{\Sigma}{\Delta} dr^2 +\Sigma{\, d\theta}^2 +\frac{ \sin ^2 \theta}{\Sigma} \left(( r^2+a^2)^2 -a^2 \Delta  \sin
   ^2\theta \right) {d\phi}^2 \, , \label{KN_metric}
   \ea
where
\be
\Sigma=r^2+a^2\cos^2\theta \, , \quad  \Delta=r^2+a^2+Q^2-2 M r\,.
\ee
and electromagnetic potential
\be
A=A_{\nu}dx^{\nu}=\frac{Qr}{\Sigma}dt+\frac{-Qar\sin^2\theta}{\Sigma}d\phi
\ee
The Hamiltonian of a particle with charge $q$ is given by 
\be
H=\frac{1}{2}g^{\mu \nu}(p_{\mu}+qA_{\mu})(p_{\nu}+qA_{\nu})  \label{H_Cparticle}
\ee
where $p_{\mu}$ is the generalized four-momentum.
Here $A_{\mu}$ is the  four-vector potential of the test particle.  For a Kerr-Newman black hole with metric~(\ref{KN_metric}),
the $t$ and $\phi$ coordinates are cyclic that
lead to two conserved quantities, namely energy and azimuthal angular momentum, with the corresponding
Noether symmetry generators $ \xi^{\mu}_{(t)}$ and $ \xi^{\mu}_{(\phi)}$, expressed as
\be
\xi^{\mu}_{(t)} = \delta^{\mu}_t \, , \quad \quad\quad  \xi^{\mu}_{(\phi)}=\delta^{\mu}_\phi \, .
\ee
 Thus, the associated  conserved energy  $\varepsilon$ and an azimuthal angular momentum  $\ell$  per unit mass for the motion of a neutral particle
 can be constructed  as
\be
 \varepsilon = - p_{\mu}\xi^{\mu}_{(t)}/m \, , \quad\quad
    \ell =  p_{\mu}\xi^{\mu}_{(\phi)}/m \, ,
   \ee
which can be realized from the effective potential to be defined later when $ r  \rightarrow \infty$.
 Using these two constants of motion,
we can find $u^t$ and $u^\phi$ given by
\be
\dot{t} = \varepsilon - \frac{(Q^2 - 2 M r)((a^2+r^2) \varepsilon - a \ell)+bQr(a^2+r^2)}{ \Delta \Sigma} \, ,
\ee
\be
\dot{\phi} = \frac{-a((Q^2-2Mr)\varepsilon+a\ell+bQr)}{\Delta \Sigma} + \frac{\ell}{\Sigma \sin^2 \theta} \,
\ee
Where $b=q/m$, the over  dot means the derivative with respect to the proper time $\tau$.
We will discuss two different kinds of Carter constant due to different purpose.
With Carter constant chosen to be

\be
\kappa^{\star}=u_{\mu}u_{\nu}K^{\mu \nu} \,
\ee

where

\be
K^{\mu \nu} = \Delta [k^{\mu}j^{\nu}-k^{\nu}j^{\mu}] + r^2 g^{\mu \nu}\ ,
\ee
\be
j^{\mu}=\frac{1}{\Delta} [(r^2+a^2)\delta^{\mu}_t + \Delta \delta^{\mu}_r + a \delta^{\mu}_\phi] \ ,
\ee
\be
k^{\mu}=\frac{1}{\Delta} [(r^2+a^2)\delta^{\mu}_t - \Delta \delta^{\mu}_r + a \delta^{\mu}_\phi] \ 
\ee
We found the equation of $\theta$-motion
\be
\Sigma^2 \dot{\theta}^2 = \kappa^{\star} -a^2 \cos^2 \theta-(a \varepsilon \sin \theta - \frac{\ell}{\sin \theta})^2 \ 
\ee
Using the fact $u^{\mu} u_{\mu} = -1$, we got equation of r-motion
\be
\Sigma^2 \dot{r} = [(r^2+a^2)\varepsilon - a \ell - bQr ]^2 - \Delta[r^2+\kappa^{\star}] \equiv R(r)^{\star} \
\ee
We than use circular motion condition $R(r)^{\star} =R'(r)^{\star} =0$ to obtain the corresponding energy and angular momentum analytically

\be
\begin{split}
\ell^{\star}(r) = \frac{r[\kappa^{\star}(2Q^2-3Mr)+(\kappa^{\star}+Q^2-Mr)r^2-a^4]+a^2[\kappa^{\star}(M+r)-r(Q^3-3Mr+r^2)]}{2ar \sqrt{\kappa^{\star}+r^2\sqrt{\Delta}}}\\
+\frac{bQ}{2ar}(a^2-r^2)  \
\end{split}
\ee

\be
\varepsilon^{\star}(r)=\frac{bQ}{2r}+\frac{\kappa^{\star}(r-M)+r(a^2+Q^2-3Mr+2r^2)}{2r \sqrt{\kappa^{\star}+r^2\sqrt{\Delta}}}
\ee
plug $\ell^{\star}(r)$ and $\varepsilon^{\star}(r)$ into $R''(r)^{\star} =0$, we got the equation of $r_{ISCO}$.
\be
6 r_{ISCO}(M-bQ\varepsilon^{\star}(r))+(6 r_{ISCO}^2+a^2)(\varepsilon^{\star}(r)^2-1)+(a\varepsilon^{\star}(r)-\ell^{\star}(r))^2+Q^2(b^2-1)-(\ell^{\star}(r)^2+\kappa^{\star})=0
\ee
this is useful when $\kappa^{\star}$ is arbitrary ~\cite{EHAC} .

It is possible to choose different Carter constant while the modified term is  combination of $\varepsilon$ and $\ell$. 
In our case, it is more convenient to choose Carter constant as
\be
\kappa^{\dagger}=u_{\mu}u_{\nu}K^{\mu \nu}-(a\varepsilon-\ell)^2 \,
\ee
the equation of $\theta$-motion here is 
\be
\Sigma^2 \dot{\theta}^2 = \kappa^{\dagger}+ (\ell-a\varepsilon)^2-a^2 \cos^2 \theta-(a \varepsilon \sin \theta - \frac{\ell}{\sin \theta})^2 \ 
\ee
and for r-motion
\be
\Sigma^2 \dot{r} = [(r^2+a^2)\varepsilon - a \ell - bQr ]^2 - \Delta[r^2+\kappa^{\dagger}+(\ell-a\varepsilon)^2] \equiv R(r) \label{rdotsquare}
\ee
Now we consider a particle orbits a rotating black hole in the equatorial plane
 by choosing $\theta=\frac{\pi}{2}$, $\dot{\theta}=0$, we found that $\kappa^{\dagger}=0$.
\be
R(r)=[(r^2+a^2)\varepsilon - a \ell - bQr ]^2 - \Delta[r^2+(\ell-a\varepsilon)^2] \label{R}
\ee

The equation~(\ref{R}) allows us to define the effective potential $V_{eff}$ by requiring
\be
\frac{1}{2} \dot{r}^2+V_{eff}(r,\alpha)=\frac{\varepsilon^2-1}{2}
\ee
where $\alpha$ denotes a collection of the parameters of the Kerr-Newman black hole and the test charged particle, namely $\alpha=(M, Q, a, \varepsilon, \ell, b)$
\be
\begin{split}
V_{eff}(r,\alpha)=\frac{-M+bQ\varepsilon}{r}+\frac{a^2-(b^2-1)Q^2+2\ell(\ell+a\varepsilon)}{2r^2}\\
-\frac{(\ell-a \varepsilon)(2M(\ell-a\varepsilon) +abQ)}{r^3}+\frac{(a^2+2Q^2)(\ell-a\varepsilon)^2}{r^4} \
\end{split} \label{VEFF}
\ee
Innermost stable circular orbit of a particle is purely a relativity effect, since eq.(~\ref{VEFF}) has modification due to relativity.
By solving $R(r)=R'(r)=0$ we got corresponding energy and angular momentum in stable circular orbit, it is difficult to solve it analytically in general, but in some limitation, it is possible to find them in simple steps.
For small $bQ$,  $(bQ)^2\approx0$, we got the analytical expression of  $\varepsilon(r)$ and $\ell(r)$ 
\be
\varepsilon(r)=\varepsilon(r)^{Neutral}+\gamma(r) bQ\,
\ee
\be
\ell(r)=\ell(r)^{Neutral}+\delta(r) bQ\ 
\ee
where
\be
\varepsilon(r)^{Neutral}=\frac{\Delta+a(\sqrt{Mr-Q^2}-a)}{r\sqrt{2Q^2+r(r-3M)+2a(Mr-Q^2)^{1/2}}} \ ,
\ee
\be
\ell(r)^{Neutral}=\frac{a(Q^2-2Mr)+(a^2+r^2)\sqrt{Mr-Q^2}}{r\sqrt{2Q^2+r(r-3M)+2a(Mr-Q^2)^{1/2}}} \ ,
\ee
\be
\gamma(r)=\frac{-4a(Q^2-Mr)+\sqrt{-Q^2+Mr}(3Q^2+r(r-4M)-a^2)}{2r[-2a(Q^2-Mr)+\sqrt{-Q^2-Mr}(2Q^2+r(r-3M))]} \ ,
\ee
\be
\delta(r)=-\frac{\sqrt{-Q^2+Mr}[-a(3Q^2+r(r-4M)-a^2)]+a^2[4Q^2+r(r-4M)]+r^2[\Delta-a^2]}{2r[-2a(Q^2-Mr)+\sqrt{-Q^2-Mr}(2Q^2+r(r-3M))]}
\ee

 $\varepsilon(r)^{Neutral}$ and $\ell(r)^{Neutral}$ are solved in our previous work. Plug $\varepsilon(r)$ and $\ell(r)$ into $R''(r)=0$ we got the equation of $r_{ISCO}$ in small $bQ$ limit.
 \be
 (6r_{ISCO}^2+a^2)(\varepsilon(r)^2-1)+6r_{ISCO}(M-bQ\varepsilon(r))-\ell(r)^2-Q^2=0 \
 \ee
 Since we have known the equation of $r_{ISCO}$ in case of neutral particle, it is interesting to see the modified term in $r_{ISCO}$ while we make the particle weakly charged. 
 that is,
 \be
 r_{ISCO}=r_{IN}+\Lambda(r_{IN}) bQ \
 \ee
 where $r_{IN}$ satisfies 
 \be
  (6r_{IN}^2+a^2)(\varepsilon^{Neutral}(r_{IN})^2-1)+6Mr_{IN}-\ell^{Neutral}(r_{IN})^2-Q^2=0 \
 \ee
 and the modified term $\Lambda(r_{IN})$
 \be
 \Lambda(r_{IN})=\frac{\Gamma\Omega(4Q^2\psi+r_{IN}(M(-3a^3+77aQ^2+\gamma(22a^2-42Q^2))+\Pi r_{IN}))}{\Upsilon(4aMQ^2(a^2-2a\Gamma-Q^2)+\Psi r_{IN}+M\Xi r_{IN}^2+(6\Gamma-7a)M^2 r_{IN^3}-\Gamma M r_{IN}^4)} \ ,
 \ee
\be
 \Gamma = \sqrt{-Q^2+M r_{IN}} \ ,
\ee 
\be
\Omega = (2(a\Gamma+Q^2)+r_{IN}(r_{IN}-3M))^{3/2} \ ,
\ee
\be
\psi = a(a^2-7Q^2)+(3Q^2-5a^2)\Gamma \ ,
\ee
\be
\Pi = -a(50M^2+20Q^2)+\Gamma (-3a^2+36M^2+17Q^2+5r_{IN}^2)+7Mr_{IN}(3a-4\Gamma) \ ,
\ee
\be
\Psi = a(-8Q^4+3M^2(3Q^2-a^2))+(8a^2M^2-4(a^2-1)Q^2)\Gamma \ ,
\ee
\be
\Upsilon = 2Q^2(\Gamma-a)+r_{IN}(2aM+(r_{IN}-3M)\Gamma) \ ,
\ee
\be
\Xi = -6aM^2+16a Q^2+3(a^2-3Q^2)\Gamma \ 
\ee

\begin{figure}
  \centering
  \includegraphics[scale=0.7]{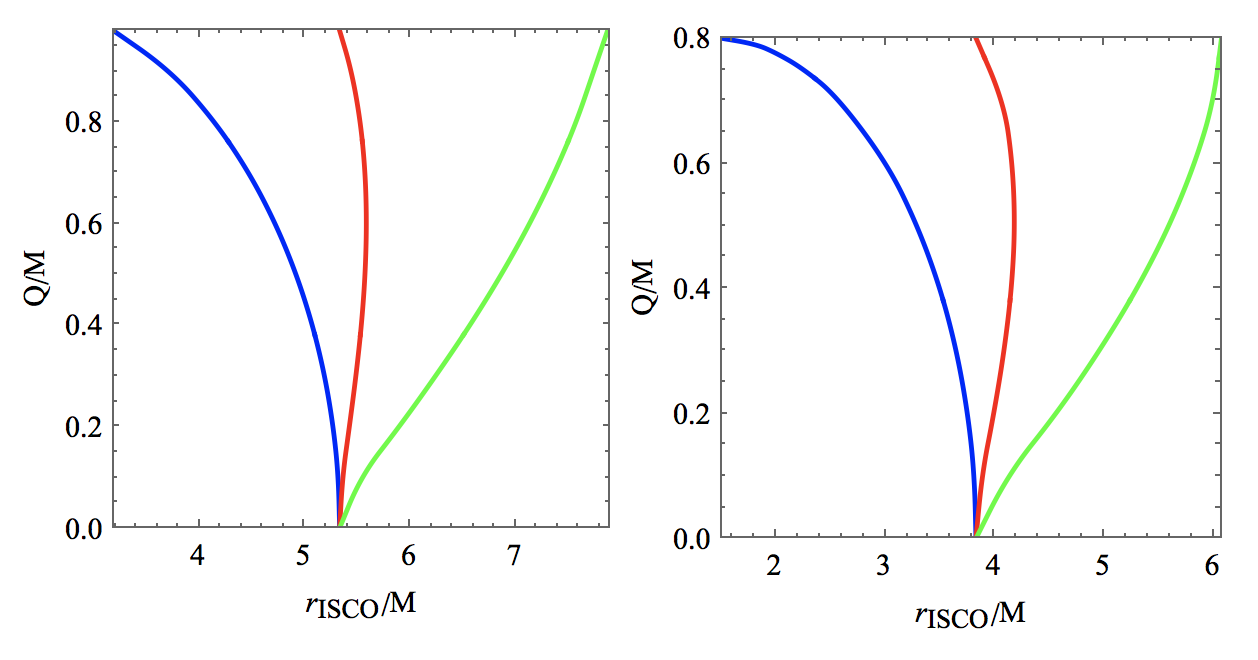}
  \caption{The dependence of the radius of the ISCO. $r_{ISCO}$ with black hole's angular momentum per unit mass $\frac{a}{M}=0.2$(left), $0.6$(right) and varing charge per unit mass $\frac{Q}{M}$. Left: Charge of particle $b=-0.1$ (blue), $-3.71815$(red, critical charge), $-10$ (green), right: $b=-0.1$ (blue), $-5.92$ (red, critical charge), $-15$(green)} \label{CCl0}
\end{figure}

\begin{figure}
  \centering
  \includegraphics[scale=0.7]{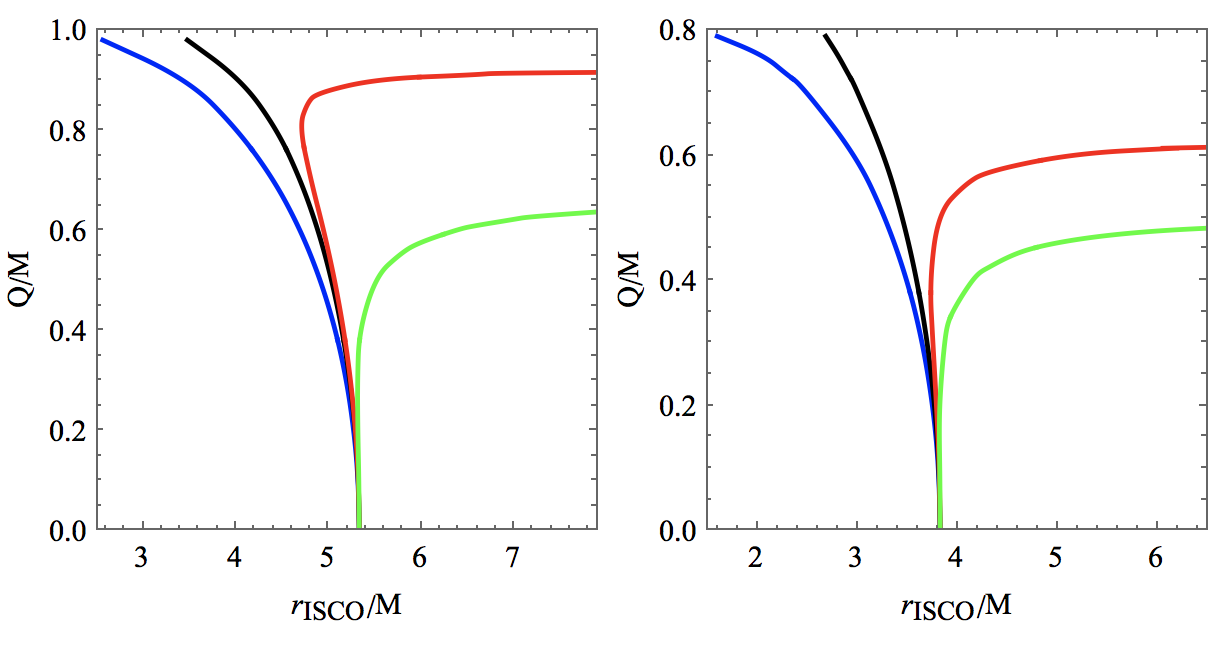}
  \caption{The dependence of the radius of the ISCO. $r_{ISCO}$ with black hole's angular momentum per unit mass $\frac{a}{M}=0.2$(left), $0.6$(right) and varing charge per unit mass $\frac{Q}{M}$. Left: Charge of particle $b=0.5$ (blue), $1.02062$(black, critical charge), $1.085$(red),$1.5$ (green), right: $b=0.5$ (blue), $1.25$ (black, critical charge), $1.6$(red),$2$(green)} \label{CCb0}
\end{figure}

\begin{figure}
  \centering
  \includegraphics[scale=0.7]{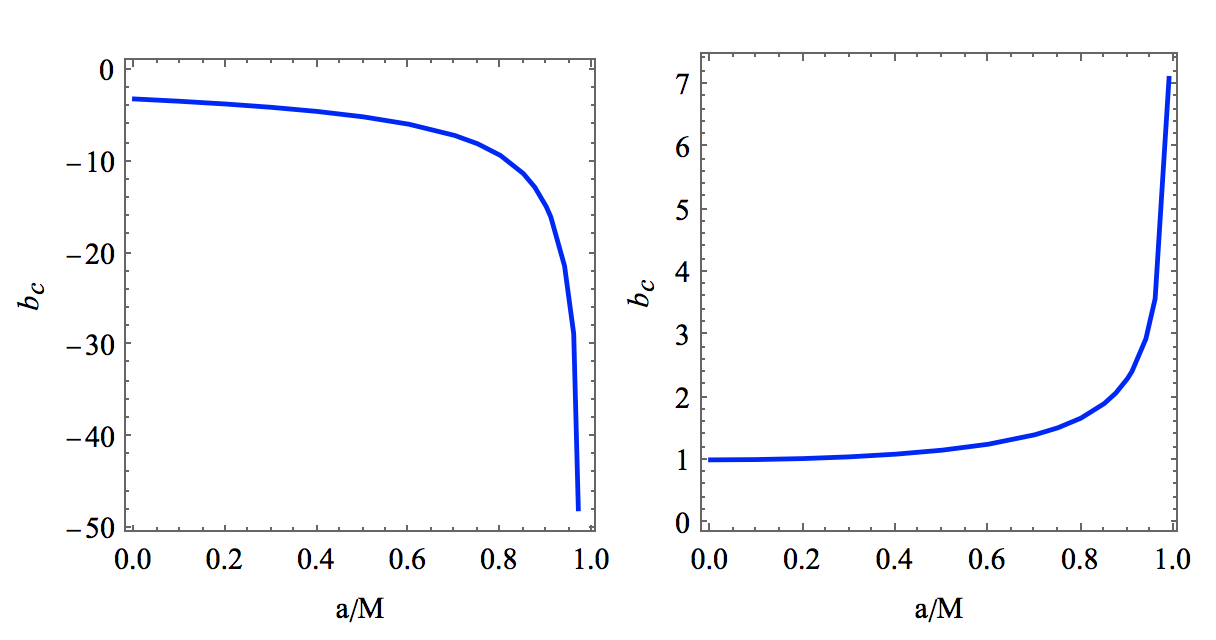}
  \caption{The dependence of the critical charge $b_{c}$ with  black hole's angular momentum per unit mass $\frac{a}{M}$ when $Q=\sqrt{M^2-a^2}$} \label{CC}
\end{figure}

 In general, we solve $R(r)=R(r)'=R(r)''=0$ for $r_{ISCO}$ numerically. Corresponding result are shown in Fig.~\ref{CCb0} and Fig.~\ref{CCl0}.
 In  both Fig.~\ref{CCl0} and Fig.~\ref{CCb0}, we see that there are two different behavior of the dependence of the radius of the $r_{ISCO}$ when $|b|$ is sufficiently larger than a critical number $|b_{c}|$. We can solve for the value of $|b_{c}|$ from the fact that the value of $r_{ISCO}(b_{c})|_{Q=0}$ is same as $r_{ISCO}(b_{c})|_{Q=\sqrt{M^2-a^2}}$ 
 \be
 r_{ISCO}(b_{c})|_{Q=0}=r_{ISCO}(b_{c})|_{Q=\sqrt{M^2-a^2}} \, \label{RR}
 \ee
 or 
 \be
 r^{\star}=r^{\dagger}
 \ee
 Left hand side of eq.(\ref{RR}) is the $r_{ISCO}$ of neutral particle around a Kerr black hole. Which is the root of
 \be
 6Mr^{\star}-r^{\star 2}+3a^2-8a\sqrt{Mr^{\star}}=0  \label{KR}
 \ee
 Right hand side of eq.(\ref{RR}) satisfies 
 \be
 (a^2-M^2)b_{c}^2+6\varepsilon r^{\dagger} \sqrt{M^2-a^2} b_{c}+M^2+\ell^2-a^2\varepsilon^2-6r^{\dagger}[M+r^{\dagger}(\varepsilon^2-1)]=0 \label{KNR}
 \ee
 Where $\varepsilon$ and $\ell$ can be obtained by solving $R(r)=R'(r)=0$. 
 After substituting eq.(\ref{KR}) into eq.(\ref{KNR}),  we got the analytic expression of $b_{c}$.
 \be
 b_{c}=\frac{-\chi\pm\sqrt{\chi^2-4\eta\iota}}{2\eta}
 \ee
 in which
 \be
 \chi=6\varepsilon r^{\star}\sqrt{M^2-a^2} \ ,
 \ee
 \be
 \eta=a^2-M^2 \ ,
 \ee
 \be
 \iota=M^2+\ell^2-a^2\varepsilon^2-6r^{\star}[M+r^{\star}(\varepsilon^2-1)]
 \ee
 Since $|b|>|b_{c}|$, the attracting force between black hole and particle is sufficiently large that makes the unstable region of the effective potential larger, which makes $r_{ISCO}$ proportional to black hole charge $Q$. Fig.~\ref{CC} shows the dependence of the critical charge bc with black hole's angular momentum.

  In Fig.~\ref{CCb0}, $bQ>0$, ISCO radius was pushed to infinity when $bQ>\frac{M}{\varepsilon}$, this makes the first term in eq.(\ref{VEFF}) change its sign, which makes the repulsive force larger than any other effect.

\section{condition for escape from a circular orbit}
A particle at a stable circular orbit of radius $r_o$ has the four-velocity
\be
\tilde{u_\nu} = (-\varepsilon_o,0,0,\ell_o) \,
\ee
To reduce the complexity of the problem we will consider a kick that gives the particle polar velocity $v_k = -r_o \dot{\theta_k}$ without changing $\ell_o$.
The space of initial condition is therefore two-dimensional:$\{r_o,\dot{\theta_k}\}$.We can express the dependence of $\varepsilon$ on $\dot{\theta_k}$ using normalization condition $u^{\mu}u_{\mu} = -1$ without neglecting $u^{\theta}$ :
\be
\begin{split}
\varepsilon = \frac{1}{r_o^4+a^2(r_o^2+2Mr_o-Q^2)}[\ell_o a(2Mr_o-Q^2)+bQr_o(r_o^2+a^2)\\
+r_o\Delta_o^{1/2}\sqrt{[r_o^4+a^2(r_o^2+2Mr_o-Q^2)](1+r_o^2 \dot{\theta_k}^2)+(r_o \ell_o-abQ)^2}\;] \, \label{energyK}
\end{split}
\ee
where $\Delta_o = \Delta|_{r=r_o}$. The root for $\varepsilon$ corresponding to future-directed four-velocity was selected.
To study the particle's behavior after the kick, it is more appropriate to reconstruct~(\ref{rdotsquare}) as
\be
\Sigma^2 \dot{r}^2 = (r^4+a^2(r^2+2Mr-Q^2))(\varepsilon-V_+)(\varepsilon-V_-) \, \label{rdotdotvpvm}
\ee
where
\be
\begin{split}
V_\pm = \frac{1}{r^4+a^2(r^2+2Mr-Q^2)}[\ell a(2Mr-Q^2)+bQr(r^2+a^2)\\
 \pm \Delta^{1/2} \sqrt{[r^4+a^2(r^2+2Mr-Q^2)](r^2+\kappa^\dagger)+r^2(r \ell-abQ)^2}\;] \, ,
\end{split}
\ee
\be
\kappa = r_o^4 \dot{\theta_k}^2 \,
\ee
$V_+(r)$ will be consider for future-directed four-velocity vector. In order to determine the escape conditions we need to inspect $V_+(r)$ to figure out how the particle moves after getting kicked.
Far away from the black hole, $V_+(r)$ becomes unity.
\be
\lim_{r \to \infty} V_{+}(r)=1
\ee
 Trivially, the particle must be energetically unbound $(\varepsilon \geq 1)$ to be able to escape. The value of $\dot{\theta_k}$ at which the particle become energetically unbound is designated as $\dot{\theta}_{\varepsilon = 1}$. We use~(\ref{energyK}) to express it as
\be
|\dot{\theta}_{\varepsilon = 1}| = [\frac{(a-\ell)[2r(a(M-bQ)-\ell M)-(a-\ell)Q^2]+r^2[2r(M-bQ)-\ell-(b^2-1)Q^2]}{r^4\Delta}]^{1/2} \,
\ee

We will assume that the trivial condition $|\dot{\theta}_k| \geq |\dot{\theta}_{\varepsilon = 1}|$ is always satisfied. When $|\dot{\theta}_k| \ll |\dot{\theta}_{\varepsilon = 1}|$, the particle oscillates slightly around the initial orbit.
The energetic freedom is not sufficient for the particle to escape when $a>0,Q>0$, in general. Depending on the black hole's parameters and particle's initial conditions, the particle may accelerate both away or toward the black hole. $V_+(r)$ has only one maximum. The particle will therefore experience only one radial turning point. Hence, the sign of the radial acceleration just after the kick $\ddot{r}(r_o)$ determines whether the particle escapes or gets captured. Using~(\ref{rdotdotvpvm}) we write an expression for $\ddot{r}(r)$ as
\be
\ddot{r}(r) = -\frac{r^4+a^2(r^2+2Mr-Q^2)}{2r^4}(\varepsilon-V_-(r))V_+'(r) \,
\ee

\begin{figure}
  \centering
  \includegraphics[scale=0.5]{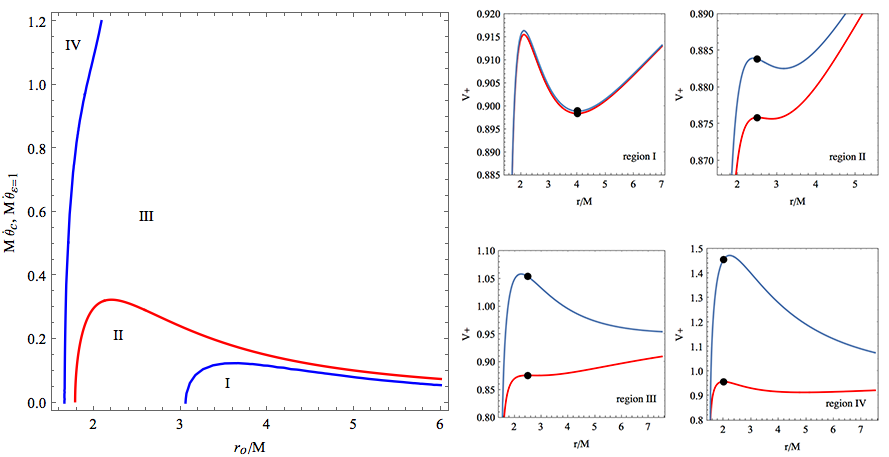}
  \caption{$|\dot{\theta_{c}}|$(blue) and $|\dot{\theta_{\varepsilon=1}}|$ vs $r_o$ for $a=0.8M$, $Q=0.3M$ and $b=0.2$} \label{ECP}
\end{figure}
\begin{figure}
  \centering
  \includegraphics[scale=0.5]{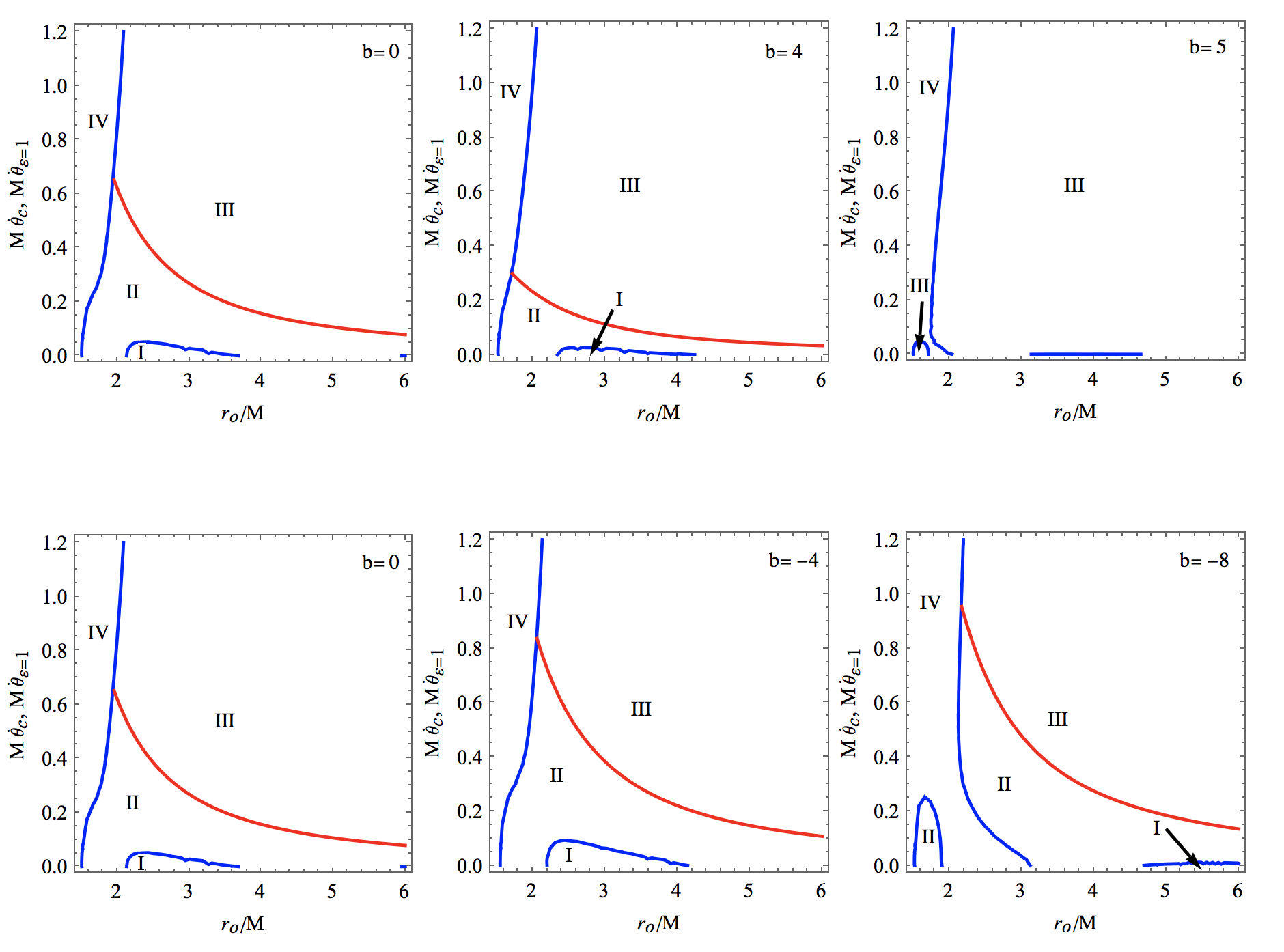}
  \caption{How charged particle affects phase with $a=0.96M$, $Q=0.2M$.} \label{CPE}
\end{figure}

Therefore, $\ddot{r}(r_o) \propto -V_+'(r_o)$ since $\varepsilon > V_-(r_o)$. Fig.~\ref{CPE} shows an example of capture, bound motion and escape.
Careful analysis of $V_+'(r_o)$ reveals that there are several distinct regions where the kicked particle accelerates in specific way.

In contrast to our previous work, a neutral particle around Kerr-Newman black hole, a charged particle is introduced.
The phase diagram in Fig.~\ref{ECP} may affected by the charge of particle. We show the behavior of "charged particle effect" in Fig.~\ref{CPE}. 

\section{Dynamics}
The dynamics of a charged particle of mass $m$, charge $q$ in curved spacetime is governed by the equation
\be
m u^{\nu} \nabla_{\nu} u^{\mu} = q F^{\mu}_{\nu} u^{\nu} \, \label{eqom}
\ee

\begin{figure}
  \centering
  \includegraphics[scale=0.7]{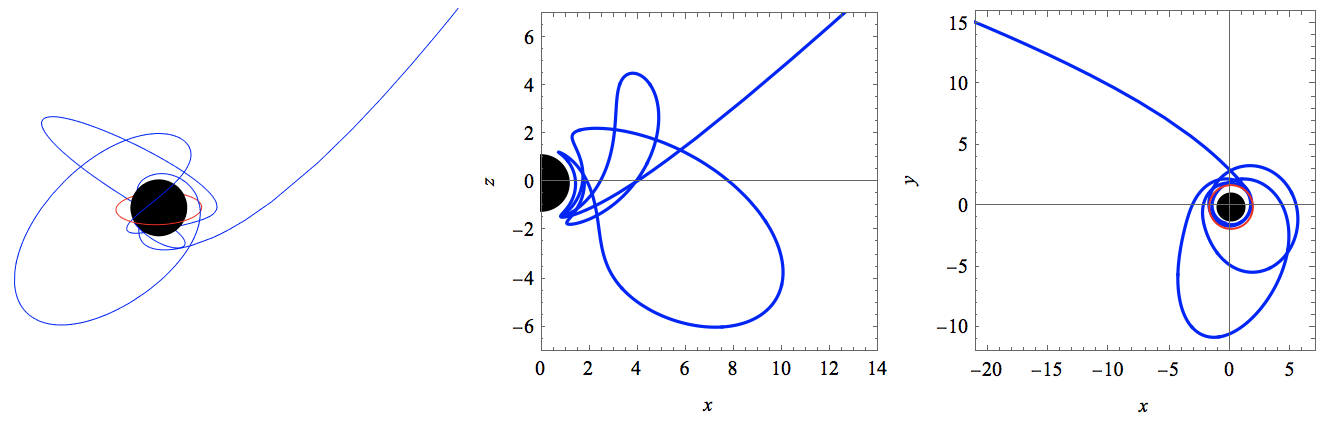}
  \caption{Escape trajectory of a charged particle kicked from circular orbit and its projection on $(x,z)$ and $(x,y)$ plane. Where $Q/M = 0.2, a/M = 0.96, r_{o}/M=1.8, M\dot{\theta}_{k}=0.37$.} \label{ET}
\end{figure}

\begin{figure}
  \centering
  \includegraphics[scale=0.7]{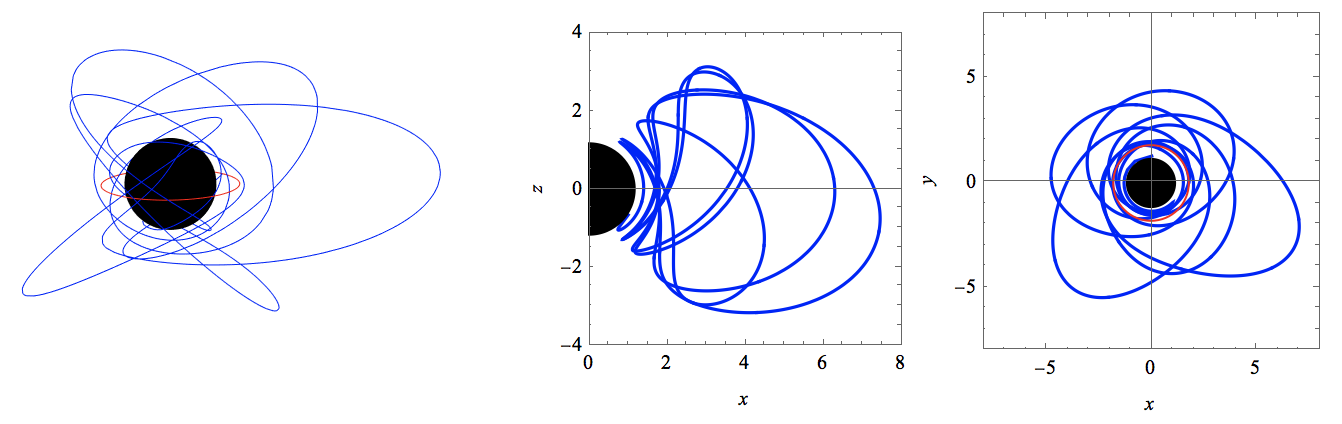}
  \caption{Captured trajectory of final states of a charged particle kicked from circular orbit and its projection on  $(x,z)$ and $(x,y)$ plane. Where $Q/M = 0.2, a/M = 0.96, r_{o}/M=1.8,  M\dot{\theta}_{k}=0.3$.} \label{CT}
\end{figure}

\begin{figure}
  \centering
  \includegraphics[scale=0.65]{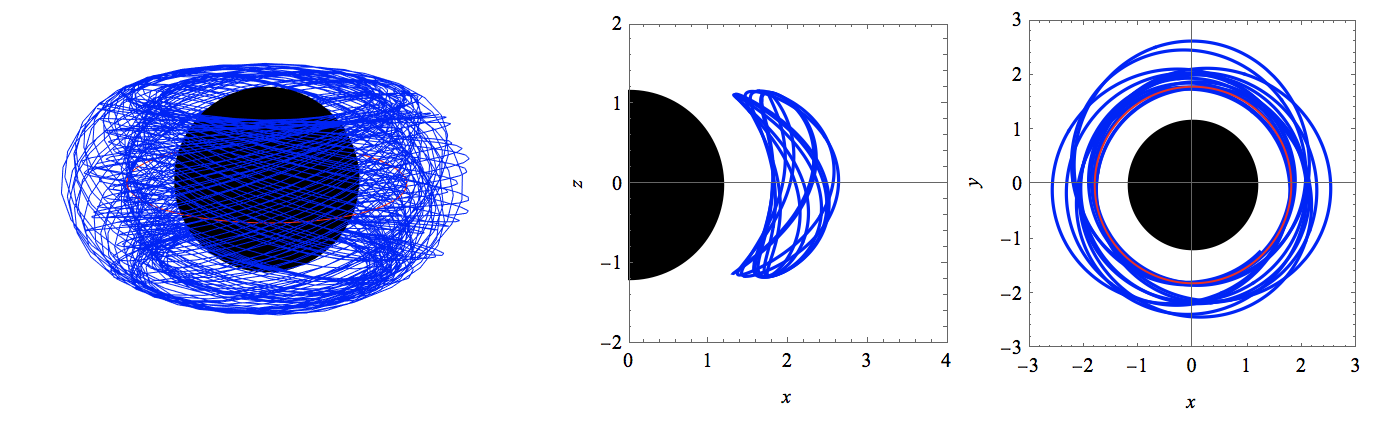}
  \caption{Bound motion trajectory of final states of a charged particle kicked from circular orbit and its projection on $(x,z)$ and $(x,y)$  plane. Where $Q/M = 0.2, a/M = 0.96, r_{o}/M=1.8,   M\dot{\theta}_{k}=0.1$.} \label{BT}
\end{figure}

The radial motion and the polar motion are obtained by solving the equation of motion~(\ref{eqom})
\be
\ddot{r} = \frac{1}{\Delta \Sigma}(\Delta^2 r \dot{\theta}^2+a^2 \Delta \sin2\theta \dot{r}\dot{\theta})+\frac{X Y}{\Delta \Sigma^5}(aZ\sin^2\theta-\frac{Y}{4})+\frac{G Z^2}{\Delta \Sigma^5}\sin^4\theta +f_{1}b^1+f_{2}b^2 \, , \label{rdotdot}
\ee
\be
\ddot{\theta} = \frac{J Y}{\Delta^2 \Sigma^5}-\frac{2r}{\Sigma} \dot{r}\dot{\theta} + \frac{H Z^2}{\Delta^2 \Sigma^5}\cos\theta \sin^5\theta + \frac{a^2 \sin2\theta}{8\Delta^2\Sigma^5}(Y^2(Q^2+2Mr)+4\Delta\Sigma^4(\Delta\dot{\theta}^2-\dot{r}^2)) +y_{1}b^1 +y_{2}b^2  \, \label{thetadotdot}
\ee
where $X, Y, Z, P, G, J, H,f_{1},f_{2},y_{1},y_{2}$ is in Appendix. We solved ~(\ref{rdotdot}),~(\ref{thetadotdot}) numerically. The final states of the particle can be characterized into three different types: bound motion, captured by black hole or escape to spatial infinity, corresponding trajectories has shown in Fig.~\ref{ET}, \ref{CT}, \ref{BT}.

\section{Weakly Magnetized Kerr-Newman Black Holes}

\begin{figure}
  \centering
  \includegraphics[scale=0.6]{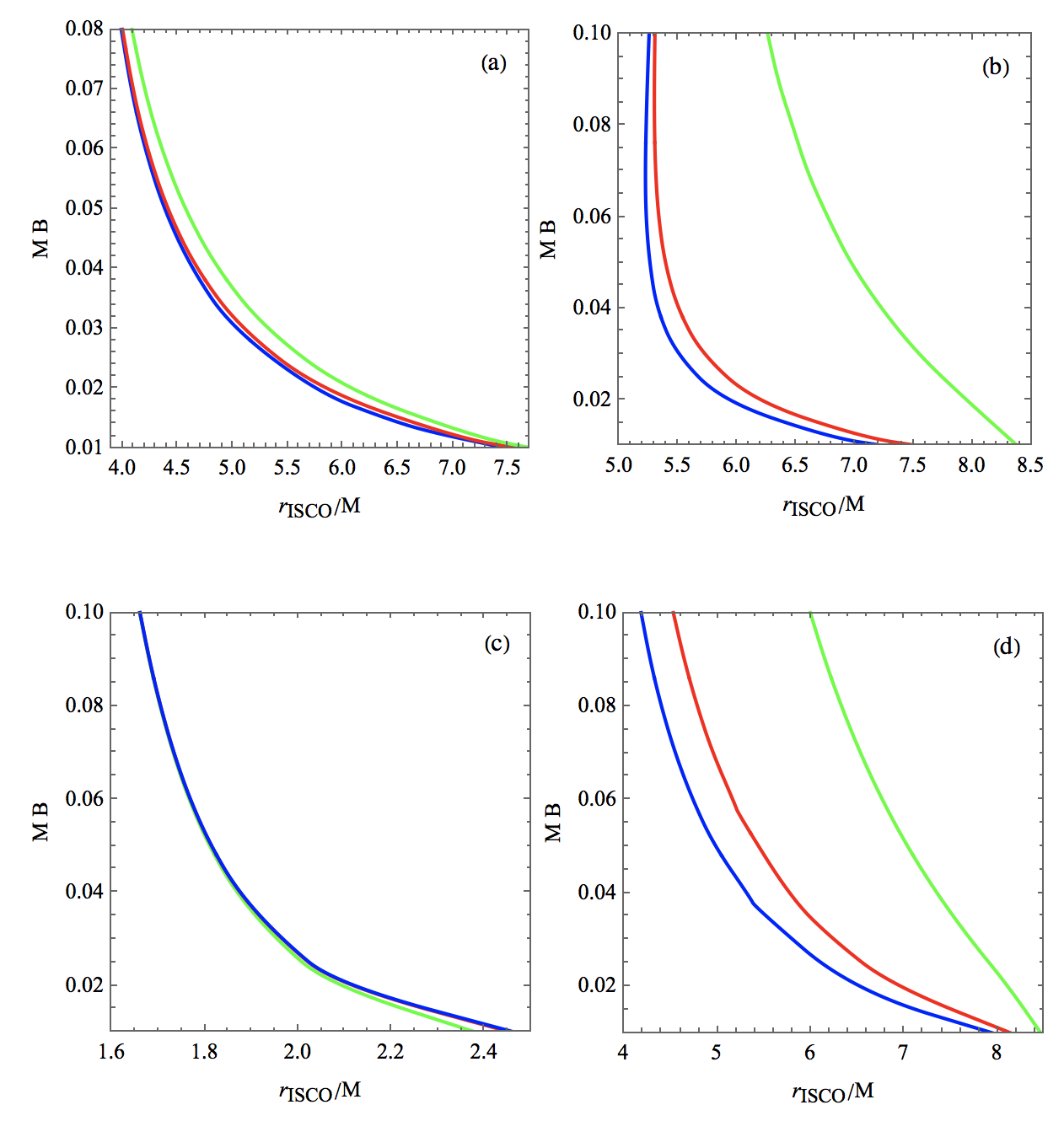}
  \caption{ISCO radius dependence on $B$ for different values of particle charge $b$ in the case of  $a/M=0.9$, $Q/M=0.4$. (a):$\ell>0$, $b=-30$(green), $-10^2$(red), $-10^6$ (blue), (b):$\ell<0$, $b=-1$(green), $-2.75$(red), $-3$(blue). (c):$\ell>0$, $b=30$(green), $10^2$(red), $10^6$(blue). (d):$\ell<0$, $b=1$(green), $5$(red), $10$(blue). }  \label{ISCOM}
\end{figure}

In this chapter, we consider the black hole is surrounded by an uniform weak magnetic field. The weak field approximation breaks down when the magnetic field creates curvature comparable to that made by black hole's mass, or 
\be
B^2\sim M^{-2}
\ee
In conventional unit, the Wald approximation fails when 
\be
B\sim \frac{k^{1/2}c^3}{G^{3/2}M}\sim 10^{49} \frac{1}{M}(Gauss)\sim 10^{26}\frac{1}{M}(meter^{-1})
\ee
Where k is the Coulomb constant. $B\sim 10^{19} Gauss\sim 10^{-4}  meter^{-1}$ for a solar mass black hole.
We set our maximum of $B$ to $10^{-1}$, means black hole mass in our case is in the scale of $10^{27} kg$, which might be 
a primordial black hole($10^{-8} kg \sim 10^{33} kg$).
The energy loss due to synchrotron radiation can be chosen to very small while we have the degree of freedom of particle charge-mass ratio $b$ ~\cite{SRAD}. 
A charged particle moves around a weakly magnetized black hole has the electromagnetic potential given by
\be
A_{\mu}=(\Phi_{0}-\Phi_{3} \omega,0,0,\Phi_{3}) \label{MVP}
\ee
$\Phi_{0}$, $\Phi_{3}$ and $\omega$ can be found at eq.(B.15) and eq.(B.17) in \cite{GWMK}, we take the maximum order of $B$ in $\Phi_{0}$ and $\Phi_{3}$  to $1$ in this paper since we set our magnetic field to be very small, which including contribution of black hole and external field.
So we have conserved quantities due to Killing vectors
\be
\varepsilon= -(g_{00}\dot{t}^2+g_{30}\dot{\phi}-bA_{0})
\ee
\be
\ell=g_{33}\dot{\phi}+g_{03}\dot{t}-bA_{3}
\ee
Using normalization condition $u^{\mu}u_{\mu}=-1$ and constrain particle in $\theta=\pi/2$ plane, $\dot{\theta}=0$, we got 
\be
\dot{t}=\frac{A_{0}bg_{33}-A_{3}bg_{30}-g_{30}\ell-g_{33}\varepsilon}{g_{03}g_{30}-g_{00}g_{33}}
\ee
\be
\dot{\phi}=\frac{A_{3}bg_{00}-A_{0}bg_{33}+g_{00}\ell+g_{03}\varepsilon}{g_{03}g_{30}-g_{00}g_{33}}
\ee
\be
\dot{r}^2=\frac{-1}{g_{11}}(1+g_{00}\dot{t}^2+g_{33}\dot{\phi}^2+2g_{30}\dot{\phi}\dot{t}) \equiv \frac{R_{M}}{r^4} \label{RM}
\ee
and
\be
\begin{aligned}
R_{M}=R_{c}+bB\{\ell[a^2(r^2-Q^2)+r^2(\Delta-a^2)] \\
+aQ[a^2(br+Q\varepsilon)+r[b(2Q^2+r(r-2M))+Qr\varepsilon]] \}\end{aligned}
\ee
$R_{c}$ is eq.(\ref{R}). With circular motion condition $R_{M}=R_{M}'=0$ and ISCO condition $R_{M}=R_{M}'=R_{M}''=0$, we solve eq.~(\ref{RM}) numerically shown in Fig.~\ref{ISCOM}. Out-going Lorentz force corresponding to anti-Larmor motion and in-going Lorentz force corresponding to Larmor motion.
Using eq.(\ref{VEFF}) we obtain the effective potential with external magnetic field modification.
\be
\begin{aligned}
V^{M}_{eff}(r)=\frac{-M+bQ\varepsilon}{r}+\frac{\ell^2+a^2(1-\varepsilon^2)+Q^2(1-b^2)}{2r^2}+\frac{(\ell-a\varepsilon)(M(\ell-a\varepsilon)+abQ)}{r^3}+\frac{Q^2(\ell-a\varepsilon)^2}{2r^4} \\
+\frac{Bb}{2}[-\ell+\frac{2\ell M+abQ}{r}-\frac{\ell(a^2+Q^2)+aQ(2bM+Q\varepsilon)}{r^2}+\frac{abQ(a^2+aQ^2)}{r^3}+\frac{a^2Q^2(\ell-a\varepsilon)}{r^4}] \label{VEFFM}
\end{aligned}
\ee
and energy-independent potential $V^{M}_{+}$

\be
V^{M}_{+}(r)=V_{0}+\frac{B(V_{10}+V_{11})}{2(a^2(Q^2-r(2M+r))-r^4)}
\ee
Where
\be
V_{s}=r^3 \Delta^{1/2}\sqrt{a^2Q^2(b^2-1)+2ar(aM+b\ell Q)+(a^2+\ell^2+r^2)r^2}
\ee
\be
V_{s1}=\sqrt{2ab\ell Qr+r^2(\ell^2+r^2)+a^2[(b^2-1)Q^2+r(2M+r)]}
\ee
\be
V_{g}=aQ\{r^2[8\ell^2+(8+b^2)r^2]+a^2[(-8+6b^2)Q^2+r(8+b^2)(2M+r)] \}
\ee
\be
V_{0}=\frac{-r^2(bQr(a^2+r^2)+a\ell(Q^2-2Mr))+V_{s}}{r^6+a^2r^2(-Q^2+r(2M+r))}
\ee
\be
V_{10}=-8aQ\frac{bQr^3(a^2+r^2)+a\ell r^2(Q^2-2Mr)-V_{s}}{r^3}
\ee
\be
V_{11}=aQ[7bQ(a^2+r^2)+8a\ell(\frac{Q^2}{r}-2M)]-\frac{\Delta^{1/2}}{V_{s1}}\{br\ell[r^4+a^2(14Q^2+r(2M+r))]+V_{g} \}
\ee
and we found that 
\be
\lim_{r \to \infty} V^{M}_{+}(r)=1+\frac{bB\ell}{2}
\ee
\begin{figure}
  \centering
  \includegraphics[scale=0.9]{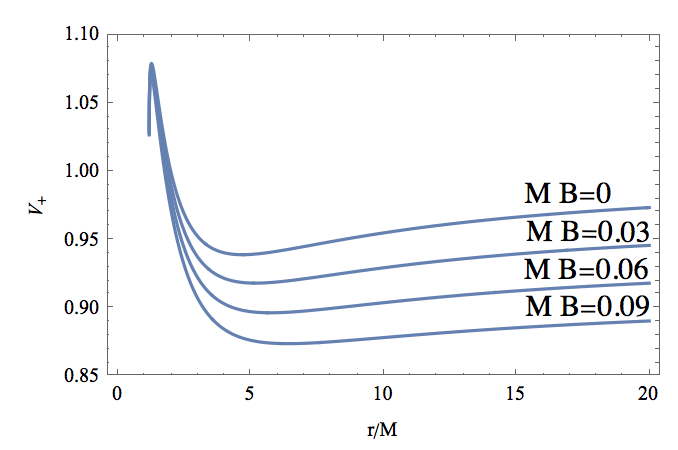}
  \caption{$V^{M}_{+}(r)$ with different $B$ and fixed $Q/M=0.4,a/M=0.9,b=-1,\ell=1.968, \varepsilon=0.918$} \label{VPMVB}
\end{figure} 
Fig.~\ref{VPMVB} shows how magnetic field $B$ affects $V^{M}_{+}(r)$. The barrier become smaller when $B$ is increased, and this lead to the decreasing value of the ISCO while $B$ is increased.

\section{Final states discussion}
\begin{figure}
  \centering
  \includegraphics[scale=0.7]{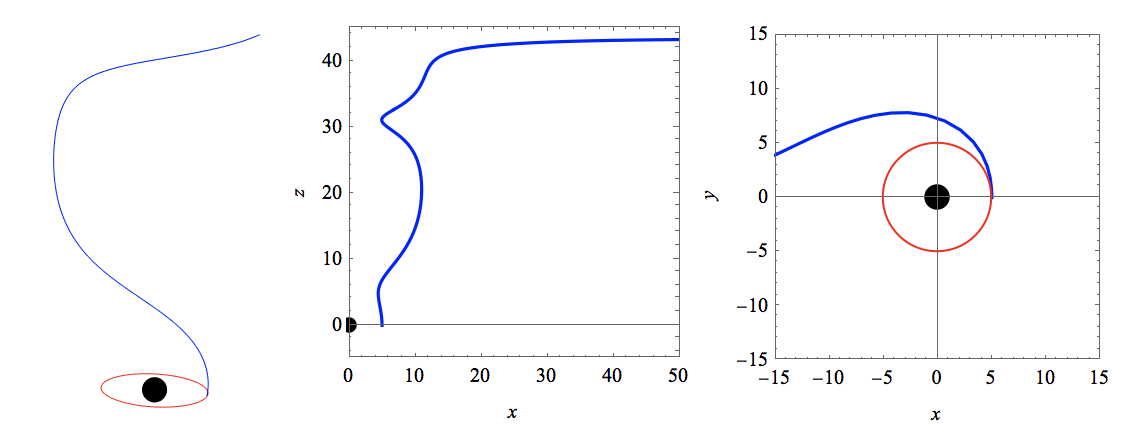}
  \caption{Escape trajectory of a charged particle kicked from circular orbit and its projection on $(x,z)$ and $(x,y)$ plane. Where $Q/M = 0.4, a/M = 0.9, r_{o}/M=5, M\dot{\theta}_{k}=0.065, M B=0.03, b=-1$.} \label{ETM}
\end{figure}

\begin{figure}
  \centering
  \includegraphics[scale=0.7]{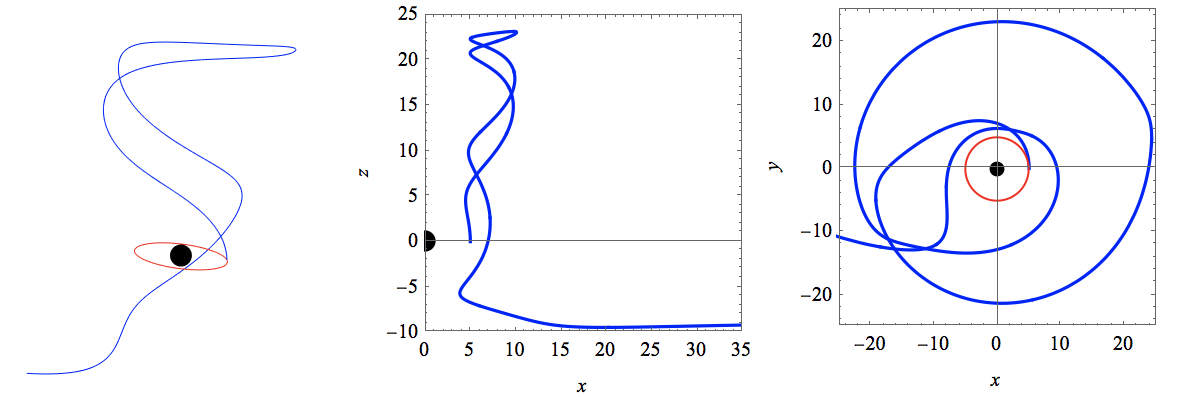}
  \caption{Escape trajectory (back scattering) of a charged particle kicked from circular orbit and its projection on $(x,z)$ and $(x,y)$ plane. Where $Q/M = 0.4, a/M = 0.9, r_{o}/M=5, M\dot{\theta}_{k}=0.06, M B=0.03, b=-1$.} \label{EBTM}
\end{figure}

\begin{figure}
  \centering
  \includegraphics[scale=0.7]{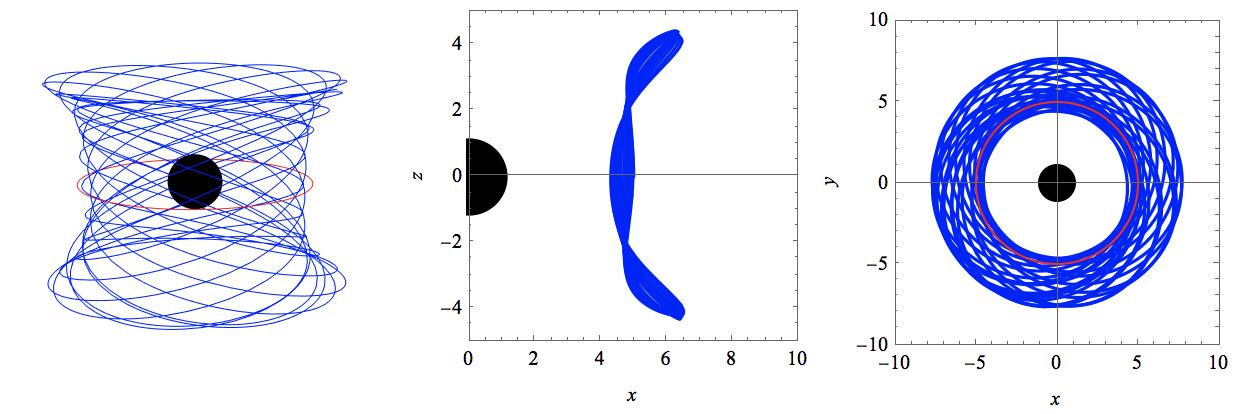}
  \caption{Bound trajectory of a charged particle kicked from circular orbit and its projection on $(x,z)$ and $(x,y)$ plane. Where $Q/M = 0.4, a/M = 0.9, r_{o}/M=5, M\dot{\theta}_{k}=0.03, M B=0.03, b=-1$.} \label{BTM1}
\end{figure}

\begin{figure}
  \centering
  \includegraphics[scale=0.7]{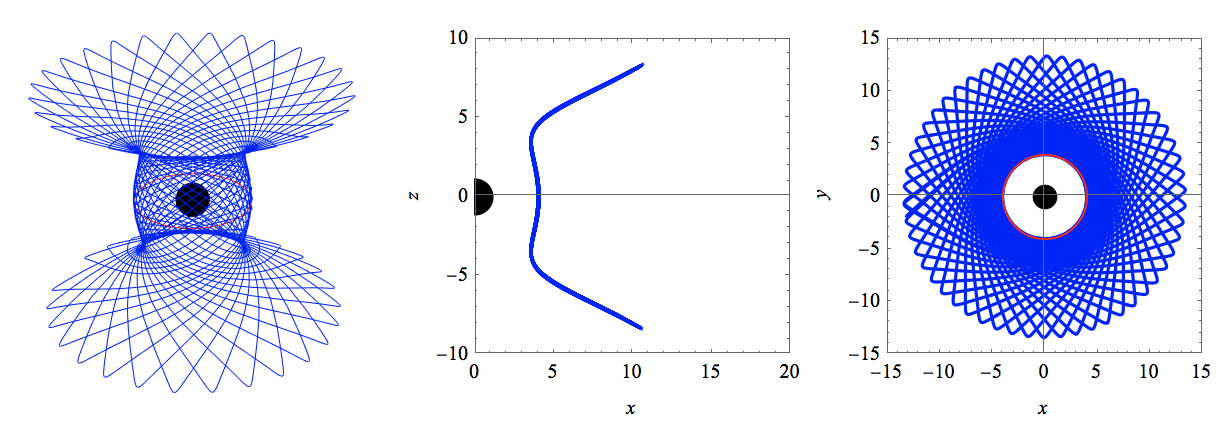}
  \caption{Bound trajectory of a charged particle kicked from circular orbit and its projection on $(x,z)$ and $(x,y)$ plane. Where $Q/M = 0.4, a/M = 0.9, r_{o}/M=4, M\dot{\theta}_{k}=0.09, M B=0.03, b=-1$.} \label{BTM2}
\end{figure}

\begin{figure}
  \centering
  \includegraphics[scale=0.7]{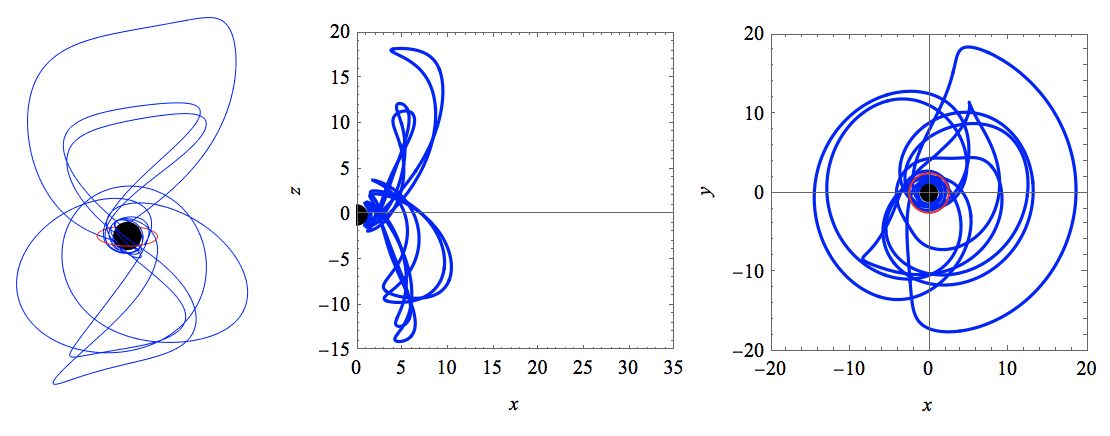}
  \caption{Captured trajectory of a charged particle kicked from circular orbit and its projection on $(x,z)$ and $(x,y)$ plane. Where $Q/M = 0.4, a/M = 0.9, r_{o}/M=2.5, M\dot{\theta}_{k}=0.29, M B=0.03, b=-1$.} \label{CTM}
\end{figure}

Since the equation of motion is non-integrable in general, to determine the final states analytically is not possible, with magnetic vector potential eq.(\ref{MVP}) substitute in eq.(\ref{eqom}), we solve the dynamic equations  numerically. Fig. \ref{ETM},\ref{EBTM},\ref{BTM1},\ref{BTM2},\ref{CTM} show some typical trajectories. 

\begin{figure}
  \centering
  \includegraphics[scale=0.7]{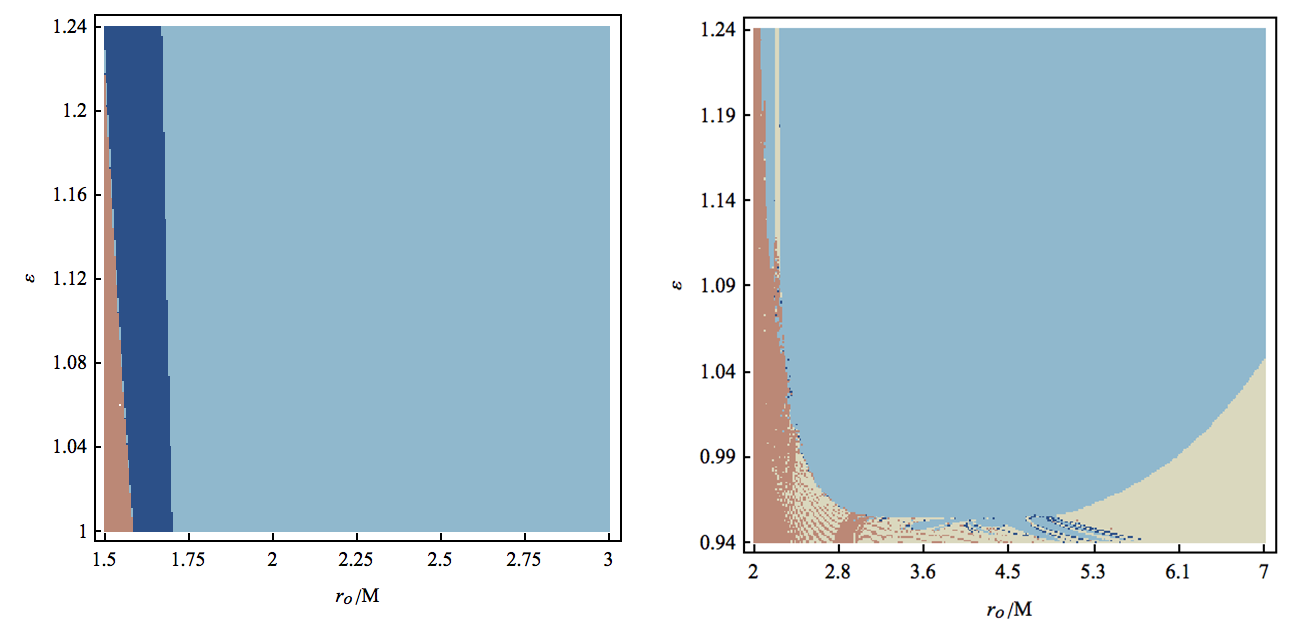}
  \caption{Chaotic behavior of basin of attraction of a charged particle kicked from circular orbit with and without an uniform magnetic field. Where $Q/M = 0.4, a/M = 0.9, M B=0 (left), 0.03(right), b=-1$.} \label{BOACEX}
\end{figure}

\begin{figure}
  \centering
  \includegraphics[scale=0.13]{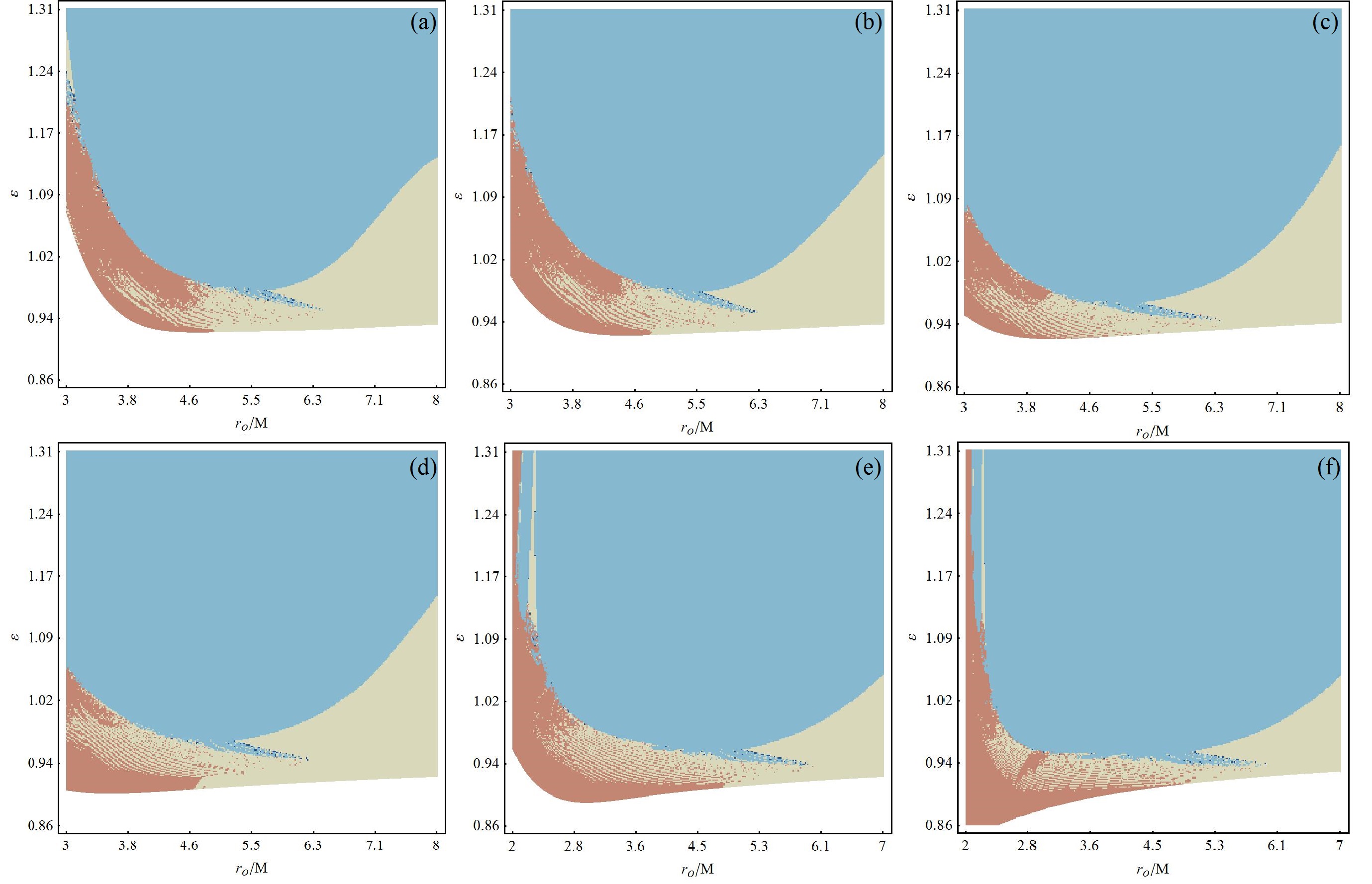}
  \caption{Basin of attraction of a charged particle kicked from circular orbit immersed in an uniform magnetic field. Where $\ell>0$(Larmor motion),$Q/M=0.4$, $M B=0.03$, $b=-1$,
  $a/M= (a) 0.15 (b) 0.3 (c) 0.45 (d) 0.6 (e) 0.75 (f) 0.9 $.} \label{LMVaA}
\end{figure}

\begin{figure}
  \centering
  \includegraphics[scale=0.13]{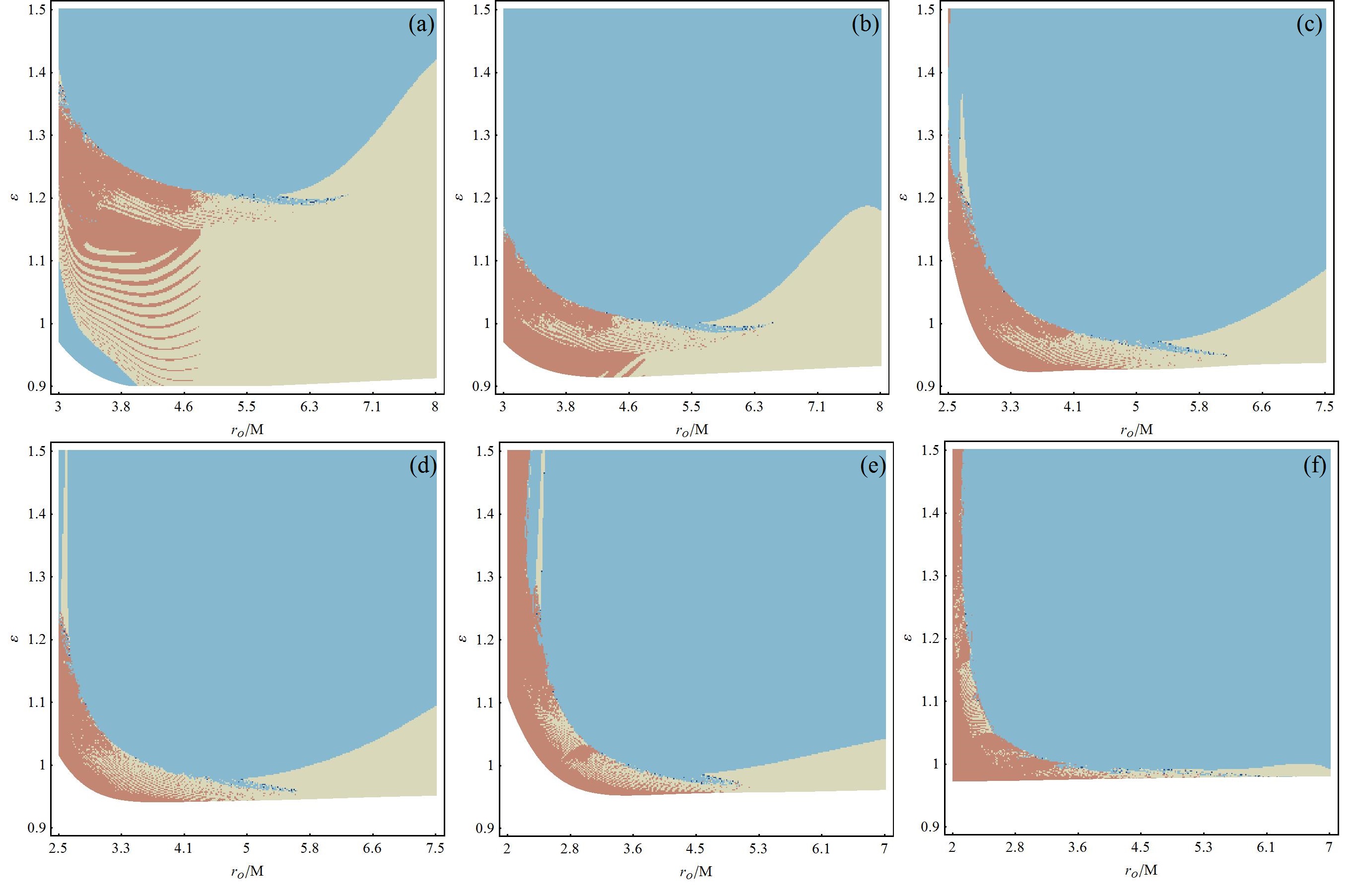}
  \caption{Basin of attraction of a charged particle kicked from circular orbit immersed in an uniform magnetic field. Where $\ell>0$(Larmor motion),$a/M=0.4$, $M B=0.03$, $b=-1$,
  $Q/M= (a) 0.15 (b) 0.3 (c) 0.45 (d) 0.6 (e) 0.75 (f) 0.9 $.} \label{LMVQA}
\end{figure}

\begin{figure}
  \centering
  \includegraphics[scale=0.13]{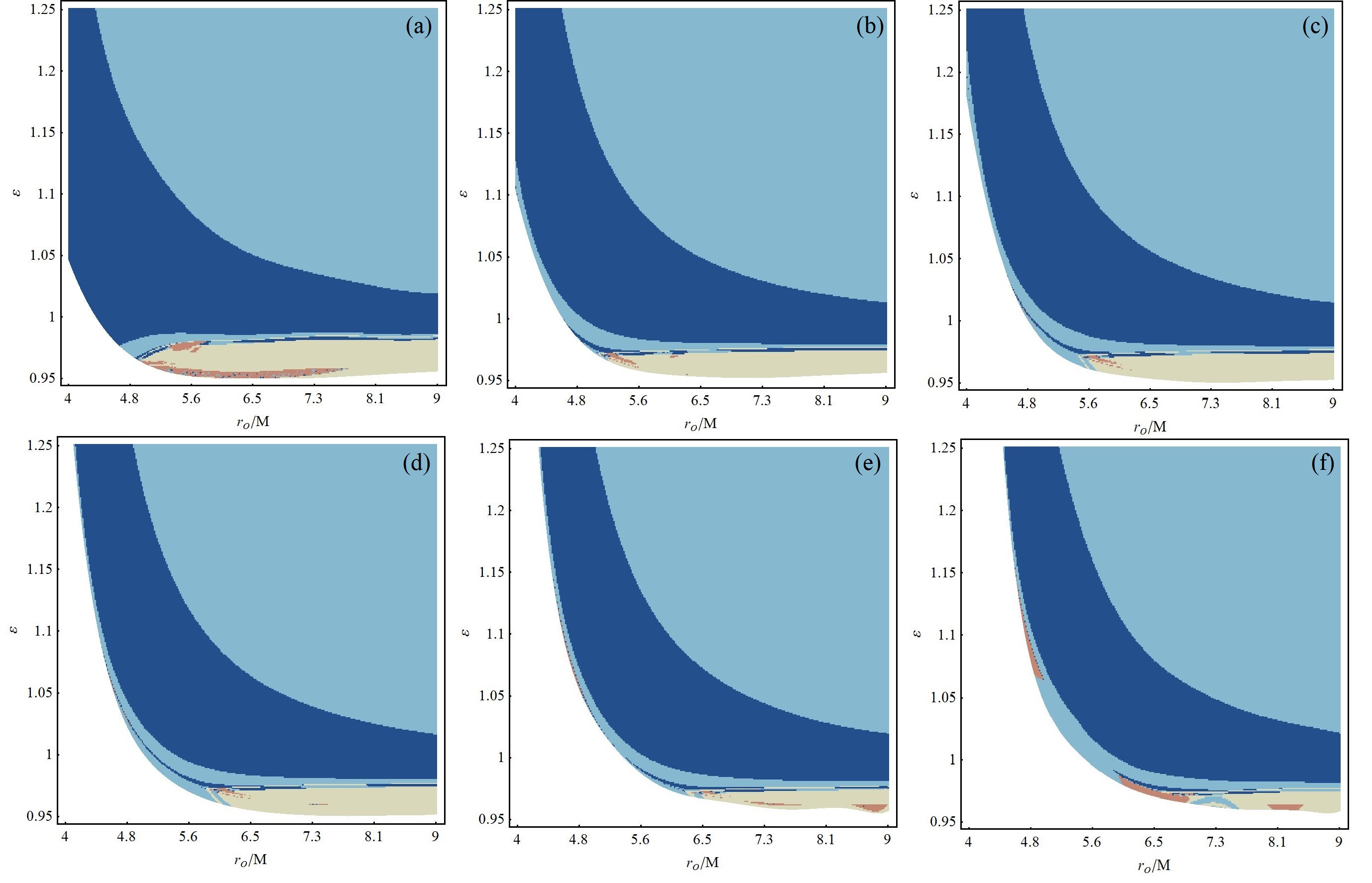}
  \caption{Basin of attraction of a charged particle kicked from circular orbit immersed in an uniform magnetic field. Where $\ell<0$(anti-Larmor motion),$Q/M=0.1$, $M B=0.01$, $b=-0.1$,
  $a/M= (a) 0.15 (b) 0.3 (c) 0.45 (d) 0.6 (e) 0.75 (f) 0.9 $.} \label{ALMVaA}
\end{figure}

\begin{figure}
  \centering
  \includegraphics[scale=0.13]{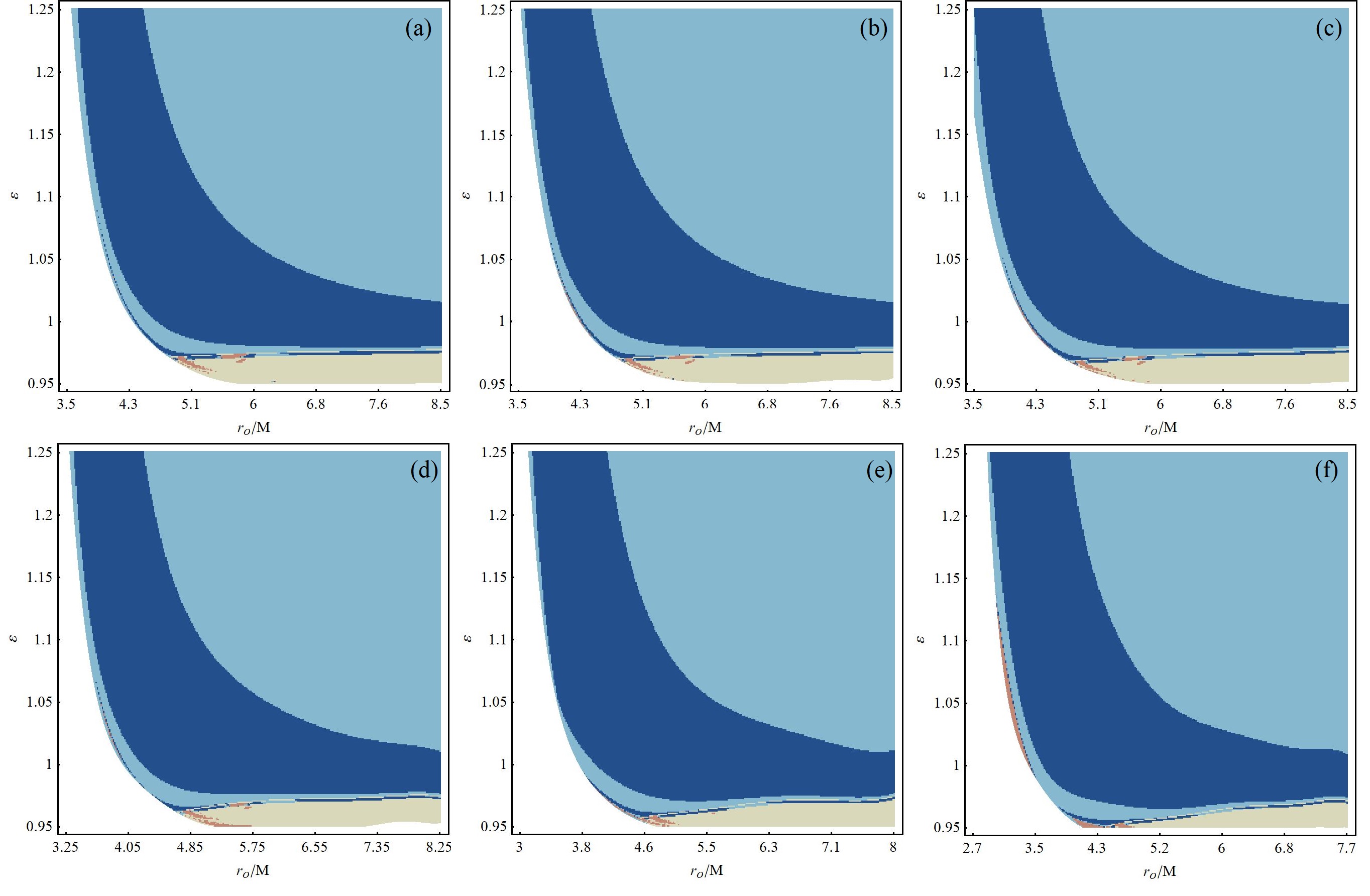}
  \caption{Basin of attraction of a charged particle kicked from circular orbit immersed in an uniform magnetic field. Where $\ell<0$(anti-Larmor motion),$a/M=0.2$, $M B=0.01$, $b=-0.1$,
  $Q/M= (a) 0.15 (b) 0.3 (c) 0.45 (d) 0.6 (e) 0.75 (f) 0.9 $.} \label{ALMVQA}
\end{figure}

An attractor of a dynamical system is a sub set of the set of all possible states of the system which an orbit with certain initial conditions approaches asymptotically. The set of initial conditions which leads to an attractor is its basin of attraction.
We set four final state with corresponding color: Escape: blue, Escape(back scattering): dark blue, Captured: red, Bound motion: pale yellow. With 2-dimension initial condition space $\{r_{o},\varepsilon\}$, we construct basin of attraction. Resolution is $300*300$ in each diagram,  Fig. \ref{BOACEX} shows chaotic behavior when an external magnetic field is introduced. Fig. \ref{LMVaA} shows the dependence of angular momentum of black hole with fixed charge of black hole and particle in case of Larmor motion, Fig. \ref{LMVQA} shows the dependence of charge of black hole with fixed angular momentum of black hole and charge of particle in case of Larmor motion,
Fig. \ref{ALMVaA} shows the dependence of angular momentum of black hole with fixed charge of black hole and particle in case of anti-Larmor motion, Fig. \ref{ALMVQA} shows the dependence of charge of black hole with fixed angular momentum of black hole and charge of particle in case of anti-Larmor motion.

\begin{figure}
  \centering
  \includegraphics[scale=0.3]{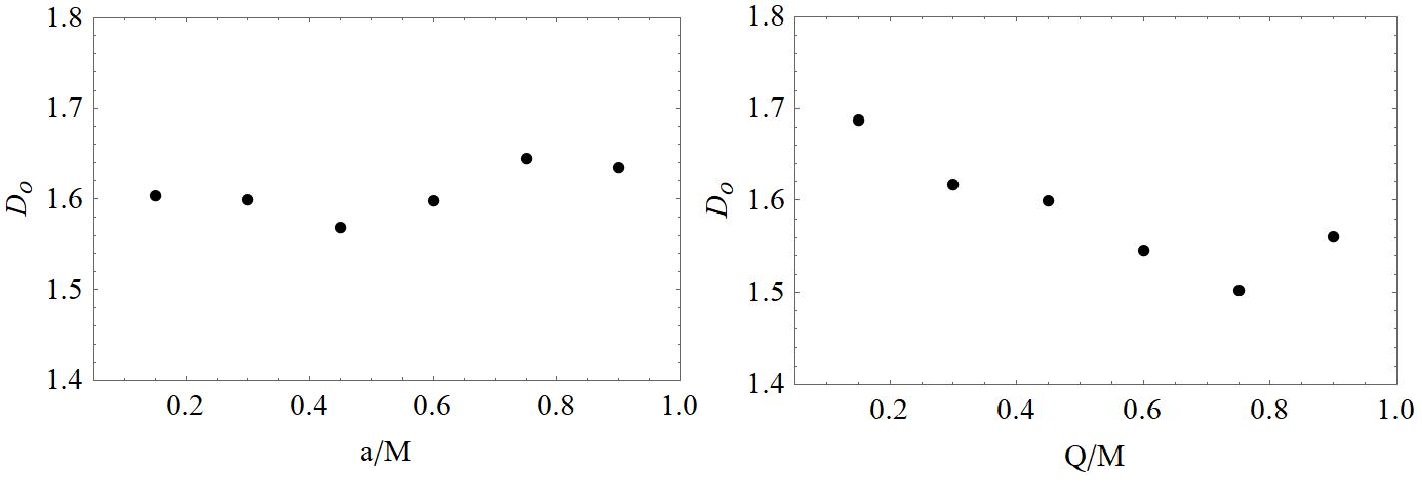}
  \caption{Fractal dimension dependence on $a/M$ and $Q/M$ in case of Larmor motion. Left: $Q/M=0.4$, right: $a/M=0.4$} \label{BCLM}
\end{figure}

\begin{figure}
  \centering
  \includegraphics[scale=0.3]{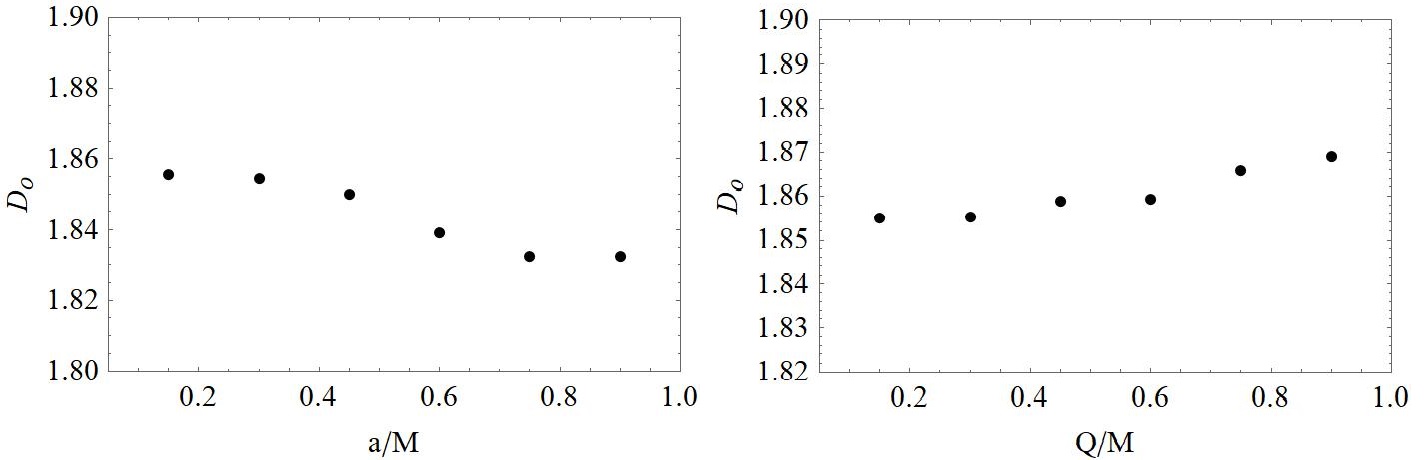}
  \caption{Fractal dimension dependence on $a/M$ and $Q/M$ in case of anti-Larmor motion. Left: $Q/M=0.1$, right: $a/M=0.2$} \label{BCALM}
\end{figure}

We use the fractal dimension $D_{o}$ of the basin boundary as a measurement of chaos. Using box-counting method one can obtain the box-counting dimension $D_{c}$
\be
D_{c}=\lim_{\epsilon \to 0} \frac{\ln N(\epsilon)}{\ln 1/ \epsilon}
\ee
for non-fractal geometrical set $D_{c}$ is an integer, if one obtain a number between integer and integer, that indicating the set is a fractal, in this case, we say the box-counting dimension $D_{c}$ the fractal dimension $D_{o}$.
Fig. \ref{BCLM} is a measurement of chaos of Fig. \ref{LMVaA} and \ref{LMVQA}, we can see that in case of Larmor motion, $D_{o}$ is nearly a constant when increasing $a/M$ and proportional to $(Q/M)^{-1}$ since the structure of bound motion and captured regions is contracted closer to event horizon while increasing $Q/M$, makes the structure simpler compared to small $Q/M$. While Fig. \ref{BCALM} is a measurement of chaos of Fig. \ref{ALMVaA} and \ref{ALMVQA}, which is anti-Laromor motion cases, $D_{o}$ decrease with increasing $a/M$ and proportional to $Q/M$ since the phase structure is away from and closer to event horizon. Different from Larmor motion cases, the phase structure roughly remain the same except for the complex region containing escape and capture trajectories. Consequently makes the different behavior of $D_{o}$ while increasing $Q/M$ in Larmor motion and anti-Larmor motion cases.

\section{Appendix}
\be
X = r(Q^2+Mr)-a^2 M\cos^2\theta \, ,
\ee
\be
Y = 2a\ell(Q^2-2Mr)+a^4\varepsilon+2r^4\varepsilon+a^2\varepsilon(2Mr+3r^2-Q^2)+a^2\Delta\varepsilon\cos2\theta \, ,
\ee
\be
Z = a(a\ell+\varepsilon(Q^2-2Mr))-\ell\Delta\csc^2\theta \, ,
\ee
\be
P = r(a^2+Q^2-Mr)+a^2(M-r)\cos^2\theta \, ,
\ee
\be
G = a^4(M-r)\cos^2\theta +r(a^2(a^2+Q^2-Mr)+2a^2(a^2+r^2)\cot^2\theta+(r^4-a^4)\csc^2\theta) \, ,
\ee
\be
\frac{J}{16 a(a^2+r^2) \cot\theta
  (Q^2-2 M r) } = 2
   \ell(\Delta
   -a^2)+a^2 \ell-a
   \varepsilon (Q^2-2 M r)+a \cos2
   \theta(a
   \ell+\varepsilon(Q^2-2
   M r)) \, ,
\ee
\be
\begin{aligned}
 H = -a^4\Delta+r^2(a^2+r^2)^2 \csc
   ^4\theta+a^2 \csc ^2\theta (a^4-r^2
  (-4 M r+2 Q^2+r^2)) \\
   +\cot
   ^2\theta(a^2 (a^2+r^2)^2
   \csc ^2\theta-2 a^4 \Delta )
\end{aligned}
\ee
\be
\begin{aligned}
f_{1}=
\frac{2r^2Q(a\ell-(a^2+r^2)\varepsilon)}{\Sigma^3}+\frac{4Q^3r^2(Q^2-2Mr)(a^2+r^2)(a\ell-(a^2+r^2)\varepsilon)}{\Delta\Sigma^5}
\\
-\frac{a\ell(a^2+Q^2+r(3r-4M))-(a^2(Q^2+2r(r-2M)+a^2)+r^2(r^2-Q^2))\varepsilon}{\Delta\Sigma^2} 
+ \frac{4Q^3r^2(a^2+r^2)\varepsilon}{\Sigma^4} 
\end{aligned}
\ee
\be
f_{2}=\frac{Q^2r(-2Q^2r^2(a^2+r^2)^2+3r^2\Delta\Sigma^2-\Sigma^3(r(M+r)+\Delta))}{\Delta\Sigma^5}
\ee
\be
\begin{aligned}
\frac{16\Delta^2\Sigma^5}{aQr}y_{1}=-64\ell\Delta^2\Sigma^2\cot\theta+
 \sin 2\theta \varpi   +a^5\Delta \sin 6\theta(a\ell+(Q^2-2Mr)\varepsilon)
 \\
  +4a^3\Delta \sin 4\theta(a^3\ell+2a\ell r^2-4Mr^3\varepsilon-a^2\varepsilon(Q^2+2Mr))
\end{aligned}
\ee
\be
y_{2}=\frac{(aQr)^2\cos\theta\sin\theta(2Q^2+(a^2+r^2)^2+\Delta\Sigma^2)}{\Delta^2\Sigma^2}
\ee
\be
\begin{aligned}
\varpi=5a^8\ell+16a^2\ell r^2(-2Q^4+r^3(r-2M)+Q^2r(4M+r)) \\
+16a^4\ell(-2Q^4+r^3(r-2M)+Q^2r(4M+r))+a^6\ell(5Q^2+r(21r-10M))-a^7(11Q^2+10Mr) \\\varepsilon-16a^3r^2\varepsilon(-2Q^4+4Mr^2(r-M)+3Q^2r(r+2M))+a^5\varepsilon(21Q^4+2M(10M-21r)r^2 \\
-Q^2r(52M+43r))+16a\varepsilon(r^4(Q^4+2M(2M-r)r^2-Q^2r(4M+r))+\Delta^2 \Sigma^2)
\end{aligned}
\ee

\end{document}